\documentclass{article}%
\begin{document}%


{\large {\bf Metric Cartesian mechanics of nonlocal energies with tensor internal tensions modifies Navier-Stokes dynamics}  
}


\bigskip
{Igor {\'E}. Bulyzhenkov} 

\smallskip
{\small {\it Moscow Institute of Physics and Technology 
 and Lebedev Physics Institute RAS} }

{\it e-mail: bulyzhenkov.ie@mipt.ru}



 \bigskip
{\bf \small Abstract}. {\small We introduce the gauge-invariant vector dynamics of continuous inertial densities through the metric formalism for extended mechanical charges. Ricci  scalar density is related to 
invariant sum of inertial and gravitational mass densities of nonlocal matter-extension. Such a Cartesian continuum of gravitating inertial densities is self-governed by internal tensor tensions toward a static equilibrium state with a Euclidean material 3-space under  the equivalence of inertial and gravitational densities of extended masses. External forces and local frictions transform  the self-dynamics of an elementary closed continuum into a forced  motion  of still adaptive energy flows, where  high-order space-time derivatives  can provide non-Newtonian self-accelerations. If  such tensor inertial feedback with the inverse constant of Cavendish  1/G is justified by measurements for the modified Navier-Stokes equation, the Newton empty space model should be  replaced by the Cartesian matter-extension for the non-local macroscopic world.}




\section{\small Introduction to elementary extended charges of Descartes}
The concept of void space regions without matter was denied logically by Des\-car\-tes. He discussed the continuous extension of matter or \lq matter-extension\rq{} as the material space plenum, early inferred by Aristotle. Newton\rq{s}  successful dynamics of point-like masses postponed the need in  sophisticated matter-extensions with the Cartesian vortex grid. Nowadays, Newton\rq{s} mechanics of localized masses and continuous gravitational fields in empty space forms the dual core of contemporary field theories. Not Cartesian physics \cite {G}, but Newtonian empty space modeling provided ontological references for 1916 Einstein\rq{s} tensor gravitation \cite {Ein}.

Dual classical physics for spatially separated matter and fields  is traditionally assigned to the macroscopic world and megacosmos. At the same time, physicists  tend to assign  nondual field densities of  quantum states only to the microscopic world. But the unique physical reality is either dual or nondual regardless of the spatial scaling and mathematical  formalisms in available theories. Similarly, the matter is either local or non-local regardless of suitable approaches to describe it.   Nonlocality of matter can be reasonably considered beyond quantum physics \cite {PopE,Eme}. The world holism \cite {Smu} and mathematical possibilities of macroscopic physics to accept the global direct overlap of extended material elements, like in quantum systems, motivates our attempts to develop the non-Newtonian world alternative with geometrization of field-like elementary charges. Our geometrization of gravitational sources will be consistent with the 1916 geometrization of gravitational fields in Einstein\rq{s} General Relativity (GR). There are no  mesoscopic scales in reality for reasonable transitions from nondual microscopic physics of continuous inertial fields to the dual macroscopic model of localized inertial masses. In our view, any mechanical state should initially be regarded as non-dual in field representations of continuous inertial energies at the submicro, micro, macro and mega-scales, at least qualitatively, if quantitative approaches have not yet been found.

It is essential for the non-empty space idea of Gustav Mie \cite{Mie} that the operator 3-density $m\delta ({\mathbf x}^\prime)$ of the point Newtonian mass in its rest-frame coordinates $\{{\mathbf x}^\prime, t^\prime \}$
can be replaced in the Poisson gravitational equation with the analytical radial density, $ m n({\mathbf x}^\prime) \equiv
 m r_{o} /4\pi |{\mathbf x}^\prime|^2 (r_o + |{\mathbf x}^\prime|)^2$, for the volume mass integral $m \equiv r_o c^2/G$ (or the extended mechanical charge $q_m \equiv {\sqrt G}m = r_o\varphi_o$) of radial inertial densities at their static gravitational equilibrium \cite {Bul,Buly}. 
The GR affine potential  $\Gamma^\mu_{\nu\mu} \equiv ln {\sqrt {-g^\prime}}=ln {\sqrt {g^\prime_{oo}}} + ln {\sqrt {\gamma^\prime}}$ reads  as $ln {\sqrt {g^\prime_{oo}}} =ln [{{r^\prime/(r^\prime +r_o)}}]$ at this static equilibrium with the  flat 3D section, $\gamma^\prime {\bf x }^\prime = 1$, in the rectangular  coordinates. Such a  radial potential of one elementary extended charge generates the  post - Coulomb/Newton fields $E_{r^\prime} \propto  \partial_{r^\prime} ln {\sqrt {g^\prime_{oo}(r^\prime)}}$, ${\mathbf \nabla}^\prime {\mathbf E}(r^\prime) \propto n(r^\prime)$ in the 3D Euclidean continuum of self-gravitating inertial densities. The possibility to unify continuous densities of radial (equilibrium) particles with their metric fields enables the unification $Q = q_m +iq_e$ of inertial and electric charges 
 in terms of complex numbers \cite {Bulyz}. 

If the imaginary densities of the combined continuous charges obey Maxwell\rq{s} equations, then the mechanical densities of electrically uncharged extended masses must also obey the same four-vector equations for the metric fields of General Relativity.
Why are energy flows in GR\rq{s} tensor gravity still different from vector currents in electrodynamics?
 The goal of this paper is to identify Maxwell-type relations between gauge-invariant tensor fields and mechanical currents. We build equal inertial and gravitational charges with zero balanced kinetic and potential self-energies.
To follow the Einstein-Hilbert geometrization of fields by the symmetrical metric tensor $g_{\mu\nu} = g_{\nu\mu}$, 
we restrict our search for asymmetrical vortex tensors for gauge invariant mechanics of extended matter by only symmetrical Christoffel\rq{s}  connections $\Gamma^\lambda_{\nu\mu} = \Gamma^\lambda_{\mu\nu}$. 
Einstein\rq{s} pure field mechanics can be fully clarified using symmetric affine connections, and below we do not consider more general gauge theories with Cartan\rq{s}  torsion.

The predominant majority  of people believe that inertial masses and electric charges are located within visible frames of  macroscopic bodies. Nevertheless, the experiment is only a criterion of truth, but not the truth itself. Nonempty space of continuous \lq matter-extension\rq{} has been recognized by philosophers not only in the Ancient East and in  Ancient Greece, but also by many  
contemporary  thinkers. \lq\lq{}A coherent field theory requires that all elements be continuous... And from this requirement arises the fact that the material particle has no place as a basic concept in a field theory. Thus, even apart from the fact that it does not include gravitation, Maxwell's theory cannot be considered as a complete theory\rq\rq{} was maintained \cite{Ei} by the author of General Relativity. Indeed, the postulated paradigm of point particles leads to the divergence of Coulomb energy, which undermines classical electrodynamics as a self-consistent theory.  A point source in the Maxwell-Lorentz equations may be considered as \lq\lq{}an attempt which we have called intellectually unsatisfying\rq\rq{} according to De Broglie \cite{Bro}. Einstein also criticized his 1915 field equation with the point gravitational source: \lq\lq{}it is	similar	to	a	building,	one	wing	of	
which	is	made	of	fine	marble	(left	part	of	the	
equation),	but	the	other	wing	of	which	is	built	of	low grade	wood	(right	side	of	equation).	The	
phenomenological	representation	of	matter	is,	in	
fact,	only	a	crude	substitute	for	a	representation	
which	would	do justice to all	known	properties	of	
matter\rq\rq{}, \lq\lq The desire ... [was] to include the gravitational field and the electromagnetic field in one unified formal picture\rq\rq{} \cite{Ein36}.


The evolution of Einstein\rq{s} theory of relativity has already passed three milestones. These are the SR postulates and the  inside energy $mc^2$ of mechanical bodies, the 1915 geometrization of massless metric fields in the Einstein Equation (based on  the Newton\rq{s} empty space dogma), and the 1938 directive to distribute particle\rq{s} mass-energy $mc^2$ continuously over all  points of the material metric space in the pure field approach to the physical reality.  Recall that the integration of particles into spatial structures of their fields was suggested by Einstein together with Infeld \cite{EInf} for the further evolution of all natural disciplines: \lq\lq We would regard matter as being made up of regions of space in which the field is extremely intense... There would be no room in this new physics for both field and matter, for the field would be the only reality.\rq\rq{} However, the extended mass  has not been yet adopted by modern relativists.  Leading authors traditionally associate Einstein\rq{s} physics only with Newtonian references and, therefore, with the Dirac delta-density of matter, the empty space metric of Schwarzschild, and the black hole model. In 1939 Einstein knowingly denied  Schwarzschild metric singularities due to the thought
 experiment \cite{Eins}. However, researchers persist in using pro-Newton delta operators to simplify the representation of elementary mass density $\mu({\bf x}) \Leftrightarrow  m \delta ({\bf x}-{\mbox{\boldmath$\xi$}})
       / {\sqrt {\gamma}}$ 
       in the classical action  $S = - c\int m ds({\mbox{\boldmath$\xi$}})$$ \Leftrightarrow - c\int\!\!\int\!\!\int\!\!\int\!\!\mu({\bf x}) ds ({\bf x}){\sqrt {\gamma}}d^3x = - c\int\!\!\int\!\!\int\!\!\int\!\!\mu({ \bf x}) (ds / {\sqrt {g_{oo}}} dx^o)  {\sqrt {\gamma g_{oo}}}  d^4x$. Delta-operator mathematics cannot describe the Cartesian 3-space $x^i$ of matter-extension, but reiterates  the Newton model of empty space with unphysical material peculiarities and the point mass path ${\mbox{\boldmath$\xi$}}(x^o)$. Below we try to consider the dynamical self-organization of elementary inertial and gravitational densities of the extended mass-energy. The  volume integral $m$  can be assigned to one Newtonian point  only for the collective fall  in external gravitational fields of the whole elementary system, $m u^\nu\nabla^{ext}_\nu u_\mu = 0$, but not for the inductive auto-shaping of  elementary mass densities $\mu^\prime = m n({\mathbf x}^\prime) $ toward the local equilibrium, $\mu^\prime u^\nu \nabla^{self}_\nu u_\mu = F^{in}_\mu $, 
			of gravitational and inertial self-forces in each nonlocal continuum.

			\section  {\small Equal equilibrium densities of inertial and gravitational charges} 
			As in the laboratory coordinates $\{ t, {\mathbf r}\}$, and in the  coordinates $\{ t^\prime, {\mathbf r}^\prime\}$ of co-moving frame,  the charge density ${\sqrt G}\mu $ $ = $ $ {\sqrt G}mn$ in the elementary inertial volume  always stays a continuous function in accordance  with  Cartesian views  on the matter-extension, the Mie\rq{s} theory of continuously charged  matter \cite{Mie},  and the Einstein-Infeld approach to purely field matter \cite{EInf}. This material spatial density can be a time-varying positive function at all points of the world energy continuum with local inertial properties. In addition, the Einstein Principle of Equivalence for inertial and gravitational masses of mechanical bodies can be coherently generalized \cite {Bul,Buly} for inertial (${\sqrt G}\mu^\prime_{in}$) and gravitational (${\sqrt G}\mu^\prime_{gr}$) charge densities of static metric continua with  local,  ${\sqrt G}\mu^\prime_{in}({\mathbf x}^\prime) \equiv {\sqrt G}\mu^\prime_{gr}({\mathbf x}^\prime)$, and nonlocal, $m_{in} \equiv \int\! \mu^\prime_{in} {\sqrt {\gamma^\prime}} d^3 x^\prime = \int\! \mu^\prime_{gr} {\sqrt {\gamma^\prime}}d^3 x^\prime \equiv   m_{gr}$,  equilibrium values. So far, the best mathematical candidate for matching the sum of paired active (inertial, kinetic) and passive (gravitational, potential) densities of relevant charges in non-equilibrium pseudo-Riemannian spacetime is the geometrical Ricci scalar $g^{\mu\nu}R_{\mu\nu} \equiv R =8\pi c^2 (	\mu_{in} + \mu_{gr})/ \varphi_o^2$. Hereinafter, $\varphi_o \equiv c^2/ {\sqrt G} = 1.04\times 10^{27} V$ is the universal self-potential for the relativistic inertial charge $q_m\equiv E_m/ \varphi_o$ related to its inside self-energy $E_m = mc^2$.
			
	Again, the equilibrium self-organization of a closed medium,  like the extended mechanical  charge \cite {Buly,BulG} with internal gravitational self-interactions, creates  a static pseudo-Riemann 4D manifold with  dilated time, $g_{oo}(r^\prime) = r^{\prime2}/(r^\prime + Gm c^{-2})^2$, and flat spatial section, $\gamma_{ij} (r^\prime) \equiv  g_{oi}g_{oj}g^{-1}_{oo} - \delta_{ij}$,  ${\sqrt {\gamma(x^\prime)}} = 1$,  ${\sqrt{-g(x^\prime)}} \equiv {\sqrt{g_{oo} \gamma} } = \sqrt{g_{oo}}$, of the elementary mass-energy system of mass $m$. Metric four-interval
	$ds^2 \equiv g_{\mu\nu}dx^\mu dx^\nu = g_{oo}(r)(dx^o)^2 - \delta_{ij}dx^idx^j$  can be applied to the equilibrium radial distribution of  matter due to six inherent symmetries $g_{oi}g_{oj} g^{-1/2}_{oo} - g_{ij} = \delta_{ij}$ for Euclidean 3-sections in rectangular coordinates. The Ricci tensor $R_{\mu\nu} = \partial_\lambda \Gamma^\lambda_{\mu\nu} - \Gamma^\lambda_{\mu\rho}\Gamma^\rho_{\nu\lambda}  + \Gamma^\lambda_{\mu\nu}\partial_\lambda ln {\sqrt {-g}} - \partial_\mu \partial_\nu ln {\sqrt {-g}}$ depends on $\sqrt{g_{oo}}$ 	for equilibrium extended matter, not on ${\sqrt{-g}}$ in arbitrarily curved non-equilibrium spaces. 
		The equilibrium spatial distribution of the elementary inertial energy $mc^2 >0$ (inside kinetic energy)  takes the spherical symmetry in the rest frame of references, where the metric components of the extended mass continuum are $g_{oo}(r^\prime,t^\prime) = 1/g^{oo}(r^\prime,t^\prime) = 1/(1 + r_o/r^\prime)^{-2} $, $g_{oi}(r^\prime,t^\prime) = g^{oi}(r^\prime,t^\prime) = 0$, $g_{ij}(r^\prime,t^\prime)= - \delta_{ij}$,  $g^{ij}(r^\prime,t^\prime)= - \delta^{ij}$, and $\partial_{t^\prime} g_{\mu\nu}(r^\prime,t^\prime) = 0$. Affine connections for equilibrium  densities of this static continuum have only two non-trivial Christoffel components, $\Gamma^i_{oo}(r^\prime,t^\prime) = \partial^i g_{oo}(r^\prime)/2$ and $\Gamma^o_{io}(r^\prime) = \partial_i g_{oo}(r^\prime,t^\prime)/2 g_{oo}(r^\prime,t^\prime)$.

The aforementioned affine connections define all components of the Ricci static tensor, including $R_{oi}(r^\prime) = 0$ and 
$R^o_o(r^\prime) = g^{oo}R_{oo} = g^{oo}(\partial_i \Gamma^i_{oo} - \Gamma^j_{oo}\Gamma^o_{oj})$ = $[ -\partial_i^2 ln (g_{oo}^{-1/2})   + (\partial_i ln (g_{oo}^{-1/2}))^2 ] = 2r_o^2 /r^2(r+r_o)^2 $. Now one can calculate the corresponding Ricci scalar,   
$R = g^{oo}R_{oo} + g^{ij}R_{ij} = g^{oo} (\partial_i \Gamma^i_{oo} - \Gamma^j_{oo}\Gamma^o_{oj})$ - $\delta^{ij}(-\partial_j \Gamma^o_{oi}- \Gamma^o_{io}\Gamma^o_{jo}) $ $= 4r_o^2 /r^2(r+r_o)^2 $ $ = 2R_o^o = - \delta^{ij}R_{ij}$ to check the cancellation of  Einstein\rq{s} curvature, $G_{oo} (x^\prime) \equiv R_{oo}(x^\prime) - g_{oo}(x^\prime) R(x^\prime)/2 =0$ and $G_{oi}(x^\prime) \equiv R_{oi}(x^\prime) - g_{oi}(x^\prime) R(x^\prime)/2 = 0$ for the static (equilibrium) elementary matter \cite{Bul, Buly}.  Geometrical self-organizations  should indeed correspond to $G_{o\mu}(x^\prime) = 0$, while  $G_{ij}(x^\prime) \neq 0$, since 3D densities of the spatial continuum of scalar masses, $\mu  \propto R \equiv - g^{\mu\nu} G_{\mu\nu} 
\Rightarrow - g^{ij} G_{ij} (x^\prime)$, cannot not depend on the non-spatial densities $G_{o\mu}(x^\prime)$.

Recall from the static states of nonlocal radial charges with ${\sqrt G} m_{in} = {\sqrt G} m_{gr}$ that their  inertial (active, kinetic) and gravitational (passive, potential) densities, respectively, define together the scalar Ricci density \cite{Bul}, 
\begin {eqnarray}
 \frac {\varphi_o R(r^\prime)}{8\pi } = \frac {\varphi_o R^o_o(r^\prime)}{4\pi  }=    \frac { {\mathbf w}^2(r^\prime)}{4\pi \varphi_o} 
- \frac { \nabla {\mathbf w}(r^\prime)}{4\pi } 
\equiv {\sqrt G}\mu_{in}(r^\prime) + {\sqrt G}\mu_{gr}(r^\prime), \label{eq1}
\end{eqnarray} 
where ${\mathbf w}(r^\prime) \equiv 
- \nabla W(r^\prime) = \Gamma^\nu_{i\nu}$, $W(r^\prime) = - \varphi_o ln (1 + r_o/r^\prime)$, $r_o \equiv {\sqrt G}m/\varphi_o$. For the co-moving frame, one can use $4\pi c^2\mu^\prime_{in} \equiv  {\mathbf w}^2(r^\prime)$ and   $ 
4\pi c^2\mu^\prime_{gr}\equiv - \varphi_o \nabla {\mathbf w}(r^\prime)$ for the static equilibrium state with equivalence 
${\sqrt G}\mu^\prime_{in} $ $  
 = {\sqrt G}m r_o/4\pi r^{\prime 2}(r^\prime+r_o)^2$ $={\sqrt G}\mu^\prime_{gr}$. From where can these continuous material densities appear in the observable  reality?	 Our next goal is  to describe, in particular, local dynamical processes in the mechanical medium beyond its metric equilibrium with $g^\prime_{oo}(r^\prime) = r^{\prime2}/(r^\prime+r_o)^2$, $g^\prime_{oi}(r^\prime)= 0$, $\gamma^\prime_{ij} \equiv g^\prime_{oi}g^\prime_{oj}(g^\prime_{oo})^{-1} - g^\prime_{ij} = \delta_{ij}$.

					
Focusing on dynamical extended charges with a possible non-equilibrium shift of inertial and gravitational densities, ${\sqrt G} m_{in} \neq {\sqrt G} m_{gr}$, we 	introduce an oscillating  prime-energy  $\varphi_o \Lambda e^{i\chi(x)}$ to describe further self-organization of resulting geometrical densities in the direction of their metric equilibrium (1) with equal inertial and gravitational charges. This elementary prime-energy of temporal oscillations with constant $\varphi_o \Lambda$ and the local phase function $\chi(x) = const + [\theta (x) +\int u_\mu(x) dx^\nu]/\Lambda$ can have the arbitrary gauge shift $\theta (x)$ next to the Feynman path integral that determines gauge-dependent prime-potentials $ U^\mu \equiv \varphi_o \Lambda\nabla^\mu e^{i\chi} \equiv i\varphi_o e^{i\chi} (u^\mu + \partial^\mu \theta) $ and  $ U_\mu \equiv \varphi_o \Lambda\nabla_\mu e^{-i\chi} \equiv -i\varphi_o e^{-i\chi} (u_\mu + \partial_\mu \theta)$. The dimensionless four-velocities $u_\mu (x) \equiv g_{\mu\nu} dx^\nu/ds$ and $u^\mu (x) \equiv dx^\mu/ds$ of material densities  in all points of the elementary space-time $x^\mu$ obey the equation of motion $u_\mu (x)u^\mu(x) = 1$.  Static mechanical 3-densities  provide the reference (1) to the Ricci scalar density. We should try to represent this Ricci scalar through compositions of vortex fields $F^{\mu\nu} \equiv (\nabla^\mu U^\nu - \nabla^\nu U^\mu)  $ and     $ F_{\mu\nu} \equiv (\nabla_\mu U_\nu - \nabla_\nu U_\mu) $.  And we tend to point  scalar densities of inertial and gravitational charges
regardless of the gauge field $\theta(x)$ and $\partial_\mu \theta(x)$.

The postulated symmetry of the Christoffel coefficients in Einstein\rq{s} metric gravitation allows to replace the covariant derivatives with partial ones in the gauge vortex fields $F_{\mu\nu}$ and $F^{\mu\nu}$, which can be represented through gauge-invariant vortex fields   $ f_{\mu\nu} \equiv (\nabla_\mu u_\nu - \nabla_\nu u_\mu) \equiv (\partial_\mu u_\nu - \partial_\nu u_\mu) $ and $ f^{\mu\nu} = g^{\mu\rho}g^{\nu\lambda}f_{\rho\lambda}$ when $(\partial_\mu\partial_\nu -\partial_\nu\partial_\mu)\chi(x) = 0$,
\begin {equation}
\cases { F_{\mu\nu} \equiv (\nabla_\mu U_\nu - \nabla_\nu U_\mu) = - i\varphi_o e^{-i\chi} f_{\mu\nu}, \
F^{\mu\nu} = g^{\mu\rho}g^{\nu\lambda}F_{\rho\lambda} = +i \varphi_o e^{+i\chi} f^{\mu\nu} \cr
U_\mu \equiv \varphi_o \Lambda \nabla_\mu e^{-i\chi} \equiv -i\varphi_o e^{-i\chi} (u_\mu + \partial_\mu \theta), \  
\chi(x)\! =\! [C \!+\!\theta (x) \!+\! \int u_\mu(x) dx^\nu]/\Lambda , \cr
4\pi J^\mu/c \equiv \nabla_\nu F^{\mu\nu} \equiv i \varphi_o e^{+i\chi}[\nabla_\nu f^{\mu\nu} + i (u_\nu + \partial_\nu \theta) f^{\mu\nu}]. 
}
 \end {equation}
   The complex four-vector $J^\mu (x)$ contains the oscillating current   $ie^{+i\chi}j_{gr}^\mu(x) \equiv ie^{+i\chi} c \varphi_o\nabla_\nu  f^{\mu\nu}/4\pi $ in the extended gravitational charge, the acceleration (orthogonal) four-vector  $ie^{+i\chi}j_{ac}^\mu(x) \equiv e^{+i\chi}\varphi_o  cu_\nu f^{\nu\mu}/4\pi $, with $u_\mu j_a^\mu \equiv 0$,  and the gauge contribution $ie^{+i\chi}j_\theta^\mu(x) \equiv  - c e^{+i\chi} \varphi_o  f^{\mu\nu} \partial_\nu \theta /4\pi $, with $ j_\theta^\mu \partial_\mu \theta \equiv 0$. 
		The four-acceleration $a^\mu \equiv  c^2u_\nu f^{\nu\mu}$ of local material densities in the nonlocal metric organization $dx_\mu = g_{\mu\nu}dx^\nu$  ($u_\mu u^\mu = 1$ and $u_\nu\nabla^\mu u^\nu = 0$)  are read  equally as $c^2u^\nu\nabla_\nu u^\mu$.

The inside  kinetic self-energy $mc^2$ or Newton\rq{s} mass $m$ can originate in this gauge  approach from the introduced tensor oscillations $F_{\mu\nu} $ and $F^{\mu\nu}$ when they form the phase-averaged superposition $\{F_{\mu\nu}F^{\mu\nu}\}_{\chi} \neq 0$ under $\{F_{\mu\nu}\}_\chi = \{F^{\mu\nu}\}_{\chi} = 0$. Indeed, the standing scalar state $F_{\mu\nu}F^{\mu\nu} =\varphi^2_o f_{\mu\nu}f^{\mu\nu} \neq 0$ of oscillating tensor densities does not depend on gauge and can describe local inertial properties of the averaged metric pattern of material densities.  Below in this article, distributed inertia in terms of slowly varying  fields $f_{\mu\nu}$ will describe the  self-organization feedback in the elementary continuum as well as the inductive feedback of elementary material densities in applied external forces. The equilibrium radial distribution of static 3D densities $\mu_{in} (r^\prime) =m_{in} r_o / 4\pi r^{\prime2}(r^\prime+r_o)^2$ of the elementary inertial mass $m_{in}$ is associated with a steady space-time vortex:
\begin {eqnarray}
\mu_{in}^\prime (x^\prime) \equiv -\frac { F_{\mu\nu}(x)F^{\mu\nu}(x)}{8\pi c^2} 
\equiv -\frac {\varphi_o^2 f_{\mu\nu}(x)f^{\mu\nu}(x)}{8\pi c^2} 
\cr 
\equiv -\frac {c^2}{8\pi G}  {  (\partial_i u_o - \partial_o u_i) (\nabla^i u^o - \nabla^o u^i)}\ \ {_x\!\Rightarrow_{x^\prime}}
 \cr 
-\frac{m_{in}}{4\pi r_o} {
 (\partial^\prime_i {\sqrt {g^\prime_{oo}}}) (\!g^{ij}\!\partial_j u^o\! +\! g^{ij} \Gamma_{jo}^o u^o\! -\! g^{oo} \Gamma_{oo}^i u^o  
 )_{x^\prime} } 
\cr =   \frac{m_{in}}{4\pi r_o}  \partial^\prime_i {\sqrt {g^\prime_{oo}(r^\prime) }} \frac {\delta^{ij} \partial^\prime_j {\sqrt {g_{oo}(r^\prime)}}   } {{\sqrt {g_{oo}(r^\prime)}}  }  
   = \frac {m_{in} r_o}{4\pi r^{\prime 2}(r^\prime+r_o)^2} 
.\label{eq2}
\end  {eqnarray}
Here we use again $\gamma_{ij} (x^\prime) \equiv (g^\prime_{oi}g^\prime_{oj}/g^\prime_{oo}) - g^\prime_{ij} = - \delta_{ij}$, $g^{ij}(x^\prime) = -\delta^{ij}$, $v_i(x^\prime) \equiv \gamma_{ij}  (x^\prime) v^j (x^\prime) = 0
$,  $u_o (x^\prime) = 1/u^o(x^\prime) = {\sqrt {g_{oo}(r^\prime)}} = r^\prime/(r^\prime+r_o)$ for equilibrium material densities in the co-moving system of references
$x^\prime = \{ ct^\prime; {\bf r}^\prime\}$.

	

The  four-flow of  inside relativistic energy $\varphi_o j_{in}^{\nu}(x) \equiv  c \varphi_o {\sqrt G}\mu_{in}^\prime(x^\prime) u^\nu (x)
 $ of inertial mass densities $\mu^\prime_{in}(x^\prime)$ or inertial charge densities ${\sqrt G}\mu^\prime_{in}(x^\prime)$ we represent through the gauge-invariant inertial four-current $j_{in}^{\nu} \equiv {\sqrt G}\mu^\prime_{in}(x^\prime) cu^\nu (x)$ with the laboratory four-velocity $c u^\nu(x)$.  The spatial transport of inertial mass-charge densities is always accompanied by the similar transport of gravitational mass-charge densities. The gravitational current density, $j_{gr}^\nu(x) \equiv c\varphi_o \nabla_\mu f^{\nu\mu}(x)/4\pi$ may not necessary coincide with the inertial current density $ j_{in}^{\nu}(x)$ for  non-equilibrium systems, when $j_{gr}^{\nu}(x)- j_{in}^{\nu}(x) \neq 0$. For   static equilibrium of  charge densities, one can use (1) in the co-moving reference frame $x^\prime = \{t^\prime; {\bf r}^\prime\}$ and 
$j_{gr}^o(x^\prime) = j_{in}^o(x^\prime) \equiv c{\sqrt G}\mu_{in}(r^\prime)u^o \equiv c\varphi_o r_o n_o(r^\prime)u^o $,
 with ${\sqrt G}m_{in} \equiv \varphi_o r_o$ and 
 \begin {eqnarray}
4\pi r_o n(r^\prime) = 4\pi r_o n(r^\prime) u^o {\sqrt {-g^\prime}} = \partial_i [{\sqrt {-g}}(\nabla^o u^i - \nabla^i u^o)]_{x^\prime} 
\cr
\Rightarrow \partial_i [{\sqrt {g_{oo}}}  (g^{oo} \Gamma^i_{oo}u^o - g^{ij}\partial_j u^o - g^{ij} \Gamma^o_{jo} u^o )]_{x^\prime}
\cr 
= \partial_i [{\sqrt {g_{oo}}}\delta^{ij} (u^og^{oo}\partial_j g_{oo} + \partial_j u^o)]_{x^\prime} %
 = \delta^{ij}\partial_i\partial_j ln {\sqrt {g_{oo}}}_{x^\prime} = r_o^2/ r^{\prime2}(r^\prime+r_o)^2.\end {eqnarray}
Static relations $j^{gr}_o(x^\prime) = j^{in}_o(x^\prime) ={\sqrt G} \mu_{in}(x^\prime) c u_o(x^\prime) = {\sqrt G} m_{in} n (r^\prime) c {\sqrt {g_{oo}(r^\prime) }} $ of the co-variant currents are consistent with their contra-variant densities due to the similar metric equalities,  
\begin {eqnarray}
4\pi r_o n(r^\prime) u_o = g^{\nu\lambda}\nabla_\lambda (\partial_o u_\nu -\partial_\nu u_o)  \Rightarrow - g^{ij}\nabla_j \partial_i   
u_o 
\cr 
= \delta^{ij} (\partial_j \partial_i u_o -\Gamma^o_{oj}\partial_i u_o) = r_o^2/r^\prime(r^\prime+r_o)^3, 
\end {eqnarray}
for the  equilibrium metric continuum with $g_{oo}(x^\prime) = r^{\prime2}/(r^\prime+r_o)^2 = 1/g^{oo}, g_{oi} = g^{oi} = 0, g_{ij} = - \delta_{ij}, g^{ij} = - \delta^{ij}, \Gamma^o_{oi} = r_o x_i/r^{\prime2}(r^\prime +r_o),  \Gamma^i_{oo} = r_o x^i/(r^\prime+r_o)^3$. 
The calculation coincidences (4)-(5) for both covariant and contrvariant densities of gauge-invariant four-currents indicates 
  that  Cartesian mechanics of  metric energy flows is based on vortex fields, similar to Maxwell\rq{s} electrodynamics.  Below we  tend to develop an analogy between relativistic gravi-inertial laws and electrodynamics, emphasizing  the universal nature of the Maxwell tensor tension for vector forces in electrically charged media and in adaptive mechanical energies of modified Navier-Stokes streams.

	\section {\small Yin-yang paired energies of metric space-time}

Despite the equivalence of inertial  and gravitational  charges in static self-organi\-za\-tions, the kinetic (yang, positive) and potential (yin, negative) self-energies take opposite integral values for equal equilibrium distributions of both extended charges. Potential (yin, immeasurable) self-energy $E_{pot} < 0 $ of self-gravitating nonlocal distributions always occurs  in  Cartesian vortex state next to kinetic (yang, measurable) self-energy $E_{kin} = m c^2 > 0$ due to  metric self-potential $W(r^\prime) \equiv \varphi_o ln {\sqrt {-g^\prime}}= \varphi_o ln {\sqrt {g^\prime_{oo}(r^\prime)}} = -\varphi_o ln [ 1 + (r_o/r^\prime)] < 0$ of the emerging gravitational charge with the positive radial density ${\sqrt G}\mu^\prime_{gr}(r^\prime) > 0$:
  \begin {equation}
E_{kin}\!+\!E_{pot} \! = \! \!\!\int\!\!\left (\!{\sqrt G}\mu^\prime_{in}\varphi_o  + {\sqrt G}\mu^\prime_{gr}\varphi_o ln{\sqrt {g^\prime_{oo}}}\right )\!\!{\sqrt {\gamma^\prime}}d^3\! x^\prime 
 = m_{in}c^2 - m_{gr} c^2 \equiv 0.\label{eq3} 
 \end {equation}  There are no inertial charges without equal gravitational ones in metric mechanical continua under the consideration.  In other words, the rest-energy integral $m_{im}c^2$ of inside kinetic oscillations  (3) is always balanced to zero by the negative potential self-energy integral in the line of Yin-Yang dialectics of Chinese thinkers. 

Cartesian matter-extension of measurable (yang, positive) kinetic energies implies counter creation of unmeasurable (yin, negative) potential self-energies. The latter diverges in the theory of Newton\rq{s} point masses.  Infinite negative self-energies were omitted from consideration in the traditional classical mechanics and electrodynamics. However, the zero energy balance also occurs for active and passive electric (imaginary) energies of the extended electron \cite{Bulyz,BulG}.  We can say that Descartes\rq{s} metaphysics of matter-extension for Nature reestablishes  the yin-yang dialectical thinking on new, quantitative level.  How can this dialectics of yin-yang paired energies modify Newton-Euler fluid  mechanics or  GR geodesic relations?

Always positive kinetic energy  was not considered by Newton separately from the interaction potentials, althogh their negative values  cannot be measured directly in practice.  Similarly, changes of  Einstein\rq{s} kinetic (positive, yang) self-energy $+mc^2$ should not be considered separately from complementary changes of potential (negative, yin) self-energies $-mc^2$. It is possible to call all potential energies passive, trying to keep an immeasurable world in a complete picture of physical reality. Passive (yin) energies do not participate in  energy exchanges with laboratory instruments. However, these negative energies and their complementary (unmeasurable) changes should contribute to the covariant formulation of relativistic laws, just as the potential energy of Newton contributes to the nonrelativistic kinetics of a point mass.
There would be no positive inertial energies of visible bodies without their negative gravitational self-energies in (5). This yin-yang cooperation is important for the stability of elementary inertial spaces, traditionally called elementary particles. There would be no change in  kinetic energies if negative ones could not resolve this temporal evolution by appropriate compensation. The issue in question is how to upgrade the classical action $\int\!{\cal L}(t)dt = -\int\! m c ds$  of the active inertial self-energy $mc^2$ in the yin-yang line of mutual compensations in  a continuous material space.

In author\rq{s} view, one should balance the classical action of the positive relativistic energy $+mc^2$ by unmeasurable negative energy from gravitational self-potentials, $S_{yy} = - \int (mc^2 - mc^2) ds = \int\![{\cal L}(x^o) -{\cal L}(x^o)] dx^o/c = 
\int [L(x) - L(x)] {\sqrt {-g}} d^4x/c \equiv 0$. 
The coordinate folding  $|\pm x^o/c| = t$  can mathematically stay behind the irreversible time arrow $dt \geq 0$ in the classical action $\int {\cal L}(t)dt$. This folding corresponds to the yin-yang physics with paired positive (measurable) and negative (unmeasurable) Lagrange densities $L(x)$ in the compensated field action $S_{yy} =0$.
Spontaneously folded 4-space with compensated densities of kinetic and potential energies in   
all  4-volume elements  ${\sqrt {-g}}d^4x \equiv {\sqrt {g_{oo}}dx^o}{\sqrt \gamma} d^3x$ describes the action integral $S_{yy}$  for yin-yang matter from  the
  \lq void nothing\rq{:}
 \begin {eqnarray}
0 = S_{yy} \equiv \!-\frac { \varphi_o^2}{16\pi c} \!\int\!   {\biggl   (   f_{\mu\nu} f^{\mu\nu}  -  f_{\mu\nu} f^{\mu\nu} \biggl )}
 {\sqrt \gamma} d^3x   {{\sqrt {g_{oo}}}dx^o} \!
\cr 
\equiv -  \frac {1}{16\pi c} \!\int\!   { \biggl   (   F_{\mu\nu} F^{\mu\nu}  -  F_{\mu\nu} F^{\mu\nu} \biggl )}
 {\sqrt \gamma} d^3x   {{\sqrt {g_{oo}}}dx^o} \! \equiv \cr 
-  \! \int\! 
\! \left[\! \frac 
 { U_\nu\! \nabla_\mu F^{\nu\mu} } {8\pi } \! -\!\frac  { F_{\mu\nu} F^{\mu\nu}}{16\pi } 
   - \frac { \nabla_\mu\!(U_\nu F^{\nu\mu})}{8\pi }  \!\right]\!\! \frac {{\sqrt {-g}}d^4x}{c}
\! 
\cr
= \!\! \int\! 
\! \left[\! \frac 
 { \varphi_o^2 (u_\nu \!+ \!\partial_\nu \theta) \nabla_\mu f^{\nu\mu} } {8\pi } \! -\!\frac  { \varphi^2_o f_{\mu\nu} f^{\mu\nu}}{16\pi } 
   - \frac { \partial_\mu \!({\sqrt {-g}}   U_\nu F^{\nu\mu})}{8\pi {\sqrt {-g}}}  \!\right]\!\! \frac {{\sqrt {-g}}d^4x}{c}	
		.
\end {eqnarray} 
Here we used that  $U_\mu J^\mu = \varphi_o u_\mu j_{gr}^\mu$ because $U_\mu U_\nu f^{\mu\nu} =  U_\nu U_\mu f^{\nu\mu} = 0.$ Gauge -dependent integrands $(\partial_\nu\theta)(\partial_\mu {\sqrt {-g}} f^{\nu\mu}) \equiv \partial_\nu 
(\theta \partial_\mu \!{\sqrt {-g}} f^{\nu\mu})$  and $\partial_\mu \!({\sqrt {-g}}U_\nu F^{\nu\mu})$ can be omitted under variations of the action due to the Gauss surface theorem.

 Our definitions of the mass densities of a metric medium in terms of local vector displacements $u^\mu \equiv dx^\mu/(g_{\lambda\nu} dx^\lambda dx^\nu)^{1/2}$  and $u_\mu u^\mu \equiv 1$  leads in the following scalar, vector, and tensor equalities,
\begin {eqnarray}
\mu_{in}^\prime\equiv-\frac { \varphi^2_o f_{\nu\lambda}f^{\nu\lambda}} {8\pi c^2 },    \mu_{gr}^\prime  \equiv   
		\frac { \varphi^2_o u_\nu  \nabla_\mu  f^{\nu\mu}}{4\pi c^2 }     
		 ,   \frac {8\pi c^2} { \varphi^2_o }  (\mu_{gr}^\prime + \mu_{in}^\prime)  = R^\prime,     u^\mu a_\mu = 0,
 \cr  
 j_{gr}^\mu
\equiv  \frac {c\varphi_o\partial_\nu ({\sqrt {-g}}  f^{\mu\nu})  }  {4\pi \sqrt {-g}},  j_{in}^\mu \equiv  
  - \frac { f_{\nu\lambda}f^{\nu\lambda}} {8\pi } c \varphi_o  u^\mu ,  a^\mu \equiv  c^2u_\nu f^{\nu\mu} = c^2 u^\nu \nabla_\nu u^\mu,  
  \cr 
 f_{\mu\nu} \equiv  \nabla_\mu u_\nu - \nabla_\nu u_\mu = \partial_\mu u_\nu - \partial_\nu u_\mu,   \nabla_{[\mu} f_{\nu\lambda]}  \equiv 0. 
     \end {eqnarray}

The motion of the distributed mass densities $\mu_{in/gr}(x) = \gamma_{{_L}\!{_F}} \mu^\prime_{in/gr} (x^\prime)$ in the laboratory frame of references $x^\mu$ preserves the Lorentz invariant  integral $m_{in/gr} = \int \mu_{in/gr} (x) {\sqrt {\gamma (x)}}  d^3x = \int \mu_{in/gr}^\prime (x^\prime){\sqrt {\gamma(x^\prime)}} d^3x^\prime$ for the elementary extended mass-charge due to the Lorentz-Fitzgerald kinematic contraction of the  volume element, ${\sqrt {\gamma (x)}}d^3x = {\sqrt {\gamma (x^\prime)}}d^3x^\prime / \gamma_{{_L}\!{_F}}$.
Local four-currents $j^\mu_{gr}$ and $j^\mu_{in}$ of invariant extended charges ${\sqrt G}m_{in/gr}$
coincide in (8) only for the static equilibrium state (1) with the flat 3-space. In the general dynamics with internal radiation exchanges of positive (yang) energies and local drag frictions,  a non-equilibrium imbalance $N^\mu(x)\neq 0$  of gravitational and inertial four-currents can be allowed,
\begin {eqnarray}
N^\mu(x) \equiv j_{gr}^\mu (x) - j^\mu_{in} (x) \equiv  \frac {c\varphi_o\nabla_\nu   f^{\mu\nu}} {4\pi } 
+ \frac 
 { \varphi_o f_{\lambda\nu} f^{\lambda\nu} } {8\pi }cu^\mu.  
\end {eqnarray}	
This non-equilibrium shift of pair dynamical states can  lead to a local violation of the Einstein Principle of Equivalence for $u_\mu  N^\mu \neq 0$, since \begin {eqnarray}
 u_\mu  N^\mu(x)  \equiv    u_\mu  (j_{gr}^\mu - j^\mu_{in} )   \equiv c{\sqrt G}(\mu_{gr} - \mu_{in} ).
\label{eq8}
\end {eqnarray}
Antisymmetry of the tensor field $f^{\mu\nu}$ over the indexes $\mu$ and $\nu$ requires strict conservation  
 of the gravitational current density, $\nabla_{\mu} j_{gr}^\mu \equiv 0$, both in equilibrium and nonequilibrium states. There are no  requirements for a conservation of inertial flows  of positive kinetic energies $\varphi_o j^\mu_{in}$ that are associated with nonequilibrium exchanges and self-shaping of the nonlocal continuum in the direction of  its metric equilibrium (1). The mechanical boson $\varphi_o N_\mu$ for kinetic or thermal energy exchanges comes from  local changes in the  active (inertial, kinetic) energy, 
\begin {eqnarray}
  \varphi_o\nabla_\mu N^\mu(x) 
\equiv  
\frac {c \varphi^2_o}{8\pi}
 \nabla_\mu \left (u^\mu  f_{\lambda\nu} f^{\lambda\nu}   \right ) = -c^3   \nabla_\mu (\mu_{in} u^\mu). 
\end {eqnarray}

A conformal transformation $g_{\mu\nu}(x) \rightarrow  {\tilde g}_{\mu\nu}(\tilde x) =  \lambda (x) g_{\mu\nu} (x)$ with an arbitrary coordinate function, $\lambda(x) \neq const$, preserves 
the invariant structures of inertial / gravitational currents and the thermal mediator $\varphi_o {\tilde N}^\mu 
= \varphi_o  {N}^\mu/ \lambda^2 $ in (8)-(11) due to $ f_{\mu\nu} = {\tilde f}_{\mu\nu}, f^{\mu\nu} = g^{\mu\rho}g^{\nu\lambda}f_{\rho\lambda} = \lambda^2   {\tilde g}^{\mu\rho}{\tilde g}^{\nu\lambda}{\tilde f}_{\rho\lambda}  =    \lambda^2 {\tilde f}^{\mu\nu}$
and ${\sqrt {-g}} = {\sqrt {-{\tilde g}}}/ \lambda^2 $.
Non-equilibrium exchanges (11) using  positive (yang) inertial energies enable conform-invariant metric of self-governed mechanical densities in any closed nonlocal system with an integral energy conservation. Re-distributions of local inertial densities do not change the positive  inertial charge of the closed metric system. 
The nonlocal system of superelectrons with collective self-organization causes electrically charged masses
to move in non-equilibrium superconductors even against fluxes of drag (normal) electrons, for instance \cite{PRB}. We discuss the \lq electromagnetic\rq{} nature of adaptive self-tensions within a nonlocal elementary continuum in the next section.

	\section {\small Inertial self-organization in the modified Navier-Stokes dynamics}

The yin-yang dialectic of nature requires a strict balance of positive (kinetic self-energy, $+\mu_{in} c^2$) and negative (potential
self-energy, $-\mu_{gr} c^2$) contributions to nonlocal configurations of metric energies.  In practice, only positive (kinetic) energies can be measured regardless of their interpretation in Newtonian, relativistic, or quantum terms. More precisely, observations are based only on the exchange of positive kinetic energies between elementary carriers. The latter cannot be localized completely in finite spatial domaines in the Cartesian mechanics of non-empty space. The Lagrange equations of motion for a presumably localized particle in small regions do not describe the self-organization of a non-local system. The classical geodesic acceleration of \lq small\rq{} probe masses in external fields  obeys the known rule  $\mu u^\nu \nabla_\nu u^\mu = 0$ for the free motion without feedback compensations. However, if the mass density $\mu^\prime$ is part of a self-gravitating nonlocal system, like the equilibrium extended mass $m$ with the static radial density $\mu^\prime (r^\prime) = m r_o/4\pi r^{\prime2} (r^\prime+r_o)^2 $ than there is no free fall for this static density $\mu^\prime$. It is immobile  in the  gravitational field of the system due to the feedback inertial force. In addition to the accelerating gravitational self-force $\mu^\prime c^2 u^\nu\nabla_\nu u^i \equiv \mu^\prime c^2 u^o \Gamma^i_{oo}u^o \equiv \mu^\prime c^2 (\partial^i g_{oo})/2 g_{oo} $, one should expect that the self-assembling  densities of the opposing inertial forces $F^i_{in}$ will control  the dynamical self-organization, $F^\mu_{in}(x) + F^\mu_{gr}(x) =  F_{in}^\mu(x) + \mu^\prime c^2 u^\nu \nabla_\nu u^\mu = 0$. How to derive an equation for finding the local inertial  self-force $F_{in}^\mu(x)$ next to the gravitational self-action $F_{gr}^\mu(x)$ in a closed gravitational system? 
 
Now we relate the inertial feedback to  combinations of high-order  derivatives $\nabla_\lambda f^{\mu\nu}$, which goes beyond the framework of the classical Lagrange formalism. Let us employ the basic equalities (8) to calculate
covariant four-vector  
$\nabla_\nu ( f^{\nu\lambda} f_{\mu\lambda}) \equiv - (4\pi /c\varphi_o) j_{gr}^\lambda f_{\mu\lambda}  + 
f^{\nu\lambda}( - \nabla_\mu f_{\lambda\nu}   - \nabla_\lambda f_{\nu\mu} ) $ and, by using $ 2 f^{\nu\lambda} \nabla_\nu f_{\mu \lambda} \equiv - f^{\nu\lambda} \nabla_\mu f_{\lambda\nu}$ and $
u^\nu \nabla_\mu u_\nu  \equiv  0$, require  the following  feedback $F^{in}_\mu$  from higher-order derivatives
\begin {equation}
F^{in}_\mu \equiv   - \frac { \varphi^2_o}{16\pi} \nabla_\nu (  4 f_{\mu\lambda} f^{\nu\lambda} - 
\delta_\mu^\nu f_{\rho\lambda} f^{\rho\lambda} )    \equiv  - \left (\mu^\prime_{in}c^2 u^\nu   + \frac {\varphi_o}{c} N^\nu   \right )f_{\nu\mu}
\equiv - F^{gr}_\mu .
\end{equation}

The field equality (12) is valid for all non-equilibrium configurations of a nonlocal material system with a  pseudo-Riemann metric
$ds^2 = g_{\mu\nu}dx^\mu dx^\nu$. We double cheak (12) for the equilibrium self-organization of gravitating densities  with a static mass-energy distribution
$\mu^\prime (r^\prime) = mc^2  r_o/4\pi r^{\prime2} (r^\prime +r_o)^2$ and $ds^2 = g_{oo}dx^2_o - \delta_{ij} dx^i dx^j$, $g_{oo} = r^2/(r^\prime+r_o)^2$, and $r_o = mc^2/\varphi_o^2$. Static equilibrium discards all time derivatives, 3-velocities and the shift (9)  of gravitational and inertial four-currents,  $N^\nu=0$. The right hand side of (12) is  $\mu^\prime_{in}c^2 u^\nu \nabla_\nu (-u_\mu) =
\mu^\prime_{in} c^2 u^o (\Gamma^o_{o\mu} u_o) = \{ 0;  \mu^\prime_{in} c^2 [x_i r_o/r^{\prime2}(r^\prime +r_o)]$ because $u_\mu = \{r^\prime/(r^\prime +r_o); 0 \}$ and $\partial_o u_\mu = 0$. The vector term $\nabla_\nu (f_{\mu\lambda} f^{\nu\lambda}) \equiv \partial_\nu (f_{\mu\lambda} f^{\nu\lambda}) 
+ \Gamma^o_{jo}f_{\mu\lambda} f^{j\lambda} - \Gamma^o_{\mu o}f_{o\lambda} f^{o\lambda} = \{0; \partial_j (f_{i o} f^{jo})  \} $ 
$= \{0; - 2[ h h^\prime + ( h^2 /r^\prime)] x_i/r^\prime \}$ can be found through the static field intensities $f_{oi} = - r_o x_i / (r^\prime+r_o)^2 r^\prime 
\equiv - h {\sqrt {g_{oo}}} x_i/r^\prime  $ and $f^{oi} =  + r_o x^i/ r^{\prime3}  \equiv + h x^i/ {\sqrt {g_{oo}}} r^\prime$. One can use $h^\prime_r + (2h_r/r) = h_r^2 $ for the amplitude of dimensionless intensity $h \equiv r_o/(r^\prime +r_o)$  to finally find for statics that    $   (-{\varphi^2_o}/{16\pi}) \nabla_\nu (  4 f_{\mu\lambda} f^{\nu\lambda} - 
\delta_\mu^\nu f_{\rho\lambda} f^{\rho\lambda} )  = \{0,  ({\varphi^2_o x_i}/{4\pi} r^\prime) [h h^\prime + (2h^2/r^\prime)] \} = 
\{ 0;  \mu^\prime_{in}(r^\prime)c^2 h x_i/r^\prime   \}   $. In other words, the static inertial self-forces  in (12) compensate the gravitational self-action in an equilibrium nonlocal system. The inertial feedback always tends to provide equilibrium distributions of self-gravita\-ting  densities of a closed elementary system.

The pro-Newtonian  postulate $c^2u^\nu\nabla_\nu u_\mu = 0$ for GR geodesic accelerations is applicable only to the \lq small\rq{} probe body in external gravitational fields, when the probe extended system can be formally approximated by one material point. 
Different elementary masses $m_n$ and $m_k$ are different metric distributions with specific 4-intervals, $ds_n \neq ds_k$. But almost equilibrium elementary extended masses within a continuous mechanical body acquire common (Euclidean) 3-geometry. Therefore an ensemble superposition of elementary densities can be described in their joint 3-space by one universal geometry. 
From the metric equality for a closed elementary system (which is only a whole Universe in the nonlocal world approach), we can model    a local balance of forces, $ \sum_k ({F_\mu^{in} + F_\mu^{gr} + F_\mu^{ext}})_{x_k} = 0$, in an open any ensemble of mechanical densities, with
					\begin {eqnarray}
					\cases {\!	- c^2\!\sum_{k=1}^K  \!   {\mu^\prime_{in}} u^\nu (\partial_\nu u_\mu\! -\! \partial_\mu u_\nu )_{x_k} 
											= - ({ \varphi^2_o} /{16\pi}) \! \sum_{k=1}^K \! \nabla_\nu  (  4 f_{\mu \lambda} f^{\nu\lambda}  \!- \! \delta_\mu^\nu f_{\rho\lambda} f^{\rho\lambda})_{x_k} \cr +		 ({\varphi_o}/c)\sum_{k=1}^K\! N^{\nu}(x_k) \!f_{\nu\mu}(x_k) + \sum_{k=1}^K F_\mu^{ext}(x_k) \cr\cr
		\!- c\rho_{in}^\prime(x^\prime)  u_{en}^\nu(x) M_{\nu\mu}(x) =  -  ({ c^2}/16\pi G) \nabla_\nu (  4 M_{\mu \lambda} M^{\nu\lambda}  \!- \! \delta_\mu^\nu M_{\rho\lambda} M^{\rho\lambda})_x \cr + F^{drag}_\mu (x) + F_\mu^{ext}(x).
				}
										\end {eqnarray}	
												
												An observer can describe the  continuous ensemble of almost equilibrium inertial relativistic  densities $c^2\mu^\prime_{in}(x_k^\prime)$ of rest-energy integrals $m_kc^2$ by collective rest-energy density $\rho_{in}^\prime(x^\prime)c^2$ with the ensemble\rq{s} four-velocity $cu_{en}^\mu(x)$, the ensemble four-acceleration $c u_{en}^\nu M_{\nu\mu}$, with $M_{\nu\mu} \equiv  c(\partial_\nu u^{en}_\mu - \partial_\mu u^{en}_\nu)$, and the ensemble inertial 4-current $\rho_{in}^\prime(x^\prime) cu_{en}^\nu(x) \equiv \sum_{k=1}^K    \mu^\prime_{in}(x^\prime_k) c u_{en}^\nu (x_k) $ in the laboratory space-time $x$. The latter contains the Euclidean 3D section, $dl^2 = \delta_{ij}dx^idx^j =  \gamma^{k}_{ij}dx_k^idx_k^j = dl_k^2$, for overlapping elementary material spaces $x^i_k$ (without 4D superimpositions of $x^\mu_k$ due to specific 4-geometries).  The external force density $F_\nu^{ext}(x)$  from  gravi-mechanical stresses,  radiation pressure, electromagnetic forces and so on  tends to destroy the self-consistent 4D geometry of elementary extended masses $m_k$ and deduce their densities from equilibrium $\mu^k_{gr} = \mu^k_{in}$. Dynamical self-organization of elementary matter restores its Euclidean 3D geometry in all external conditions.
External stresses destroy  not only the original metric tensor $g^k_{\mu\nu}(x_k)$ of each elementary space-time $x^\mu_k$, but also the ensemble self-organization  in the direction of joint static equilibrium.   
We relate thermal inertial exchanges $\varphi_o \sum_{k=1}^K  N^{\nu}(x_k) f_{\mu\nu}(x_k)/c $  with the conventional force of drag frictions  $F_\mu^{drag} = F^{drag}_\mu (\eta, \xi)$, described by the phenomenological viscosity coefficients $\eta$ and $\xi$.

						The ensemble equation (13) is not a metric  equality, similar (12) and (8) for one closed metric system, but a model approximation of the collective motion of joint inertial density under the action of external forces.  These external forces, together with internal self-forces and  dissipative frictions determine the laboratory acceleration of the local energy density.  The feedback through  tensor tensions in (13) can modernize adaptive energy flows both in ideal Euler fluids and Navier-Stokes viscous media.  In the absence of external gravitational fields, pseudo-Euclidean 4-interval can be used for the ensemble four-velocity $cu^{en}_\mu \Rightarrow \{c, - V_i \}/{\sqrt {1- c^{-2}V^2}} $, $ cu^{en}_\mu cu_{en}^\mu = c^2,
		\delta_{ij}cu_{en}^j = - cu^{en}_i =  V_i(x)/{\sqrt {1-c^{-2} V^2(x) }}, V_i = \delta_{ij}V^j$.
		We can traditionally model internal frictions   		
		 by	the phenomenological 3-forces $F_i^{drag} (\eta, \xi)  \Rightarrow $ $ \partial_j [\eta (\partial^j V_i + \partial_i V^j)  - (2\eta/3)\delta_i^j \partial_j V^j  ]
				+ \partial_i (\xi \partial_j V^j)$ for the low speed limit $V^2/c^2 \rightarrow 0$.
		In the four-vector relations (13)		there are only 3 independent equations  due to one scalar bound $u_{en}^\mu u_{en}^\nu M_{\mu\nu} = 0$. These three equations take the following nonrelativistic form for slow turbulent flows and steady vortex patterns  of adaptive energies with inertial self-organization, 
				\begin {eqnarray}
	{\rho^\prime}\left [ \partial_t  {V_i}  + \partial_i \left (\frac {V^2}{2} \right )  - V^j M_{ji}  \right ] 
		= -	\frac { c	}{4\pi G}\partial_t (M_{i j} M^{o j} )- \frac { c^2	}{4\pi G}\partial_m (  M_{i j} M^{m j} )  \cr + \frac { c^2	}{16\pi G}\partial_i (M_{jo} M^{jo} + M_{jm} M^{jm})  
				 + \eta \rho^\prime \partial_j \partial^j V_i  + \left (\xi + \frac {\eta}{3} \right )\rho^\prime \partial_i \partial_j V^j + F_i^{ext}.
		 			 			\end {eqnarray}			
						Here we put the inertial feedback,  ${c^2} \partial_\nu ( \delta_\mu^\nu M_{\rho\lambda} M^{\rho\lambda} - 4 M_{\mu \lambda} M^{\nu\lambda}  )/16\pi G$, of tensor tensions with the Cavendish inverse constant next to the densities  of all applied forces, $F_i^{ext}$, and viscose resistance to the forced flow. New \lq electromagnetic\rq{} tensions cannot be described by textbook constructions $\partial_\nu (u^{en}_\mu u_{en}^\nu - \delta_\mu^\nu) p $ with one scalar field $p(x)$, called pressure in pro-Newtonian thermodynamics
						(toward chaos without structural self-organizations).    Our tensor tensions are well defined by one non-relativistic  velocity $V_i$  in the anti-symmetrical tensor combinations  $cM_{io}(x) \approx \partial_t V_i + \partial_i ( V^2/2)$, 
										$M_{ij} (x) \approx \partial_j V_i -  
				 \partial_i V_j$, 
				$M^{io} = - \delta^{ij}M_{jo}$,
				$M^{ij} = \delta^{im}\delta^{jl}M_{ml} $.	Such kinematic tensions of nonlocal matter do generate internal self-pressure to compensate partially the Newtonian gradients of external pressure along the moving flows, $F^{ext}_{i} \Rightarrow -\partial_i p^{ext}     \parallel  \partial_i (M_{\mu\nu} M^{\mu\nu}) $. But tensor tensions can also generate non-Newtonian self-forces $ - c^2	\partial_m (  M_{i j} M^{m j} )/{4\pi G  }$ with normal components to locally directed vector  $F^{ext}_{i}$. Note that the inertial force is proportional to $ G^{-1} $, while the force of gravity is proportional to $G^{+1}$.  	
				The relativistic model  (13)-(14) can be applied to stationary and nonstationary configurations of  \lq smart\rq{} self-assembling densities in various external conditions. 				
		By closing,  the ideal Euler fluid with scalar pressure gradients and Navier-Stokes modifications only by drag resistance cannot, in principle,  describe  new tensor tensions for large scale turbulent structures and other self-organizations in moving media.	
						
			\section {\small Conclusion for purely field theory and measurements}
			Quantum probabilities do not have a clear ontology and cannot remove outdated  Newtonian references from Einstein\rq{s} relativistic mechanics.  Cartesian metaphysics for matter-extension  criticizes Newton\rq{s} empty space and can insist on purely field terms for the sources. The field theory of matter,  inferred by Mie \cite {Mie} even before General Relativity, anticipated continuous densities of sources instead of point masses. The Yin-Yang dialectical heritage offers to create  measurable inertial energies together with immeasurable negative compensations that maintain zero balanced inertial and gravitational energies (6) in the \lq void nothing\rq{} of imaginary oscillations (2).

			Tensor gravitational fields and vector gravitational currents in (8) have the well known electromagnetic structures but with opposite singes of sources in accordance with the Poisson equation for Newtonian gravity. Representing electricity in imaginary numbers of the complex gravi-electric charge $Q \equiv q_m + i q_e = {\sqrt G}m_o +ie_o $ of the electron, one can unify  the Maxwell-type metric equalities (8) with electrodynamics of imaginary charges \cite {Bulyz, BulG}. Such Cartesian electrodynamics is free from singularities and has the correct sign of the radiation reaction force ${\bf F}_{rad} =
			2(ie_o)^2 {\dot {\bf a}} / 3 c^3$ due to $i^2 = -1$. The nonlocal nature of the extended densities in (8) equally applies to electrodynamics  of extended electrons, which  releave the nondual field theory  from the Thompson  4/3 problem \cite {34}.      
			
			In turn, the nonlocal electrodynamics of continuous elementary  charges becomes very beneficial for the metric gravi-inertia (7)-(8) of extended masses. 			There is a clear analogy of our tensor self-tensions in (13)-(14) with  the well-studied tensor tensions of electromagnetic fields.	In particular, temporary chan\-ges in the Umov - Poyting inertial transfer, $ c \partial_t (M_{i j} M^{o j} ) / 4\pi G $, and orthogonal (to main streams of energy) flows in (14) cannot not be proposed  for adaptive hydrodynamics without analogies with electrodynamics.    	
			From (13) - (14),	it should be noted that the Cavendish constant $ G $ determines not only gravity, but also the inertia of adaptive mechanical fields. If self-accelerations with the \lq inverse Cavendish constant\rq{} in the new modification (14) of  Navier-Stokes dynamic were confirmed in direct laboratory experiments with vortexes, then Cartesian physics of nonlocal material space should replace the Newtonian empty space paradigm for (non-existent) point masses.
				
				The introduced \lq electromagnetic\rq{} model for Cartesian fluid dynamics (14) allows for turbulence and large-scale vortex self-organizations of adaptive inertial flows with low viscosity coefficients. A dynamical ordering of chaotic states is in a line of the new relativistic principle \cite {B} for each probe body $k$ to move in external fields toward  equipartition of the inside chaos-energy,  $m_kc^2 {\sqrt {1-\beta^2} }$, and translational order-energy, $m_kc^2 \beta^2 / {\sqrt {1-\beta^2} }$. Toroid flows have an extra rotational degree of freedom and, due to this principle \lq toward chaos-order equipartition of kinetic self-energies\rq{,} one or few toroidal formations  can spontaneously replace an initially cylindrical vortex in liquids and gases, depending on boundary conditions.				
				Newtonian physics requires  the monotonous temporal disintegration of ordered systems into chaotic states, which denies our \lq electromagnetic\rq{}  increments for large scale self-organization.   Cartesian dynamics (14) with nonlocal  feedback of tensor self-tensions is awaiting to be falsified /justified  in time-varying rotations of media with low viscosity, including superfluid Bose condensates. Such principle experiments can measure the \lq inverse Cavendich constant\rq{} and distinguish between Cartesian and Newtonian alternatives for engineering, physics, life sciences and other academic disciplines.

	\end {document}